\documentclass[12pt]{article}
\usepackage[centertags]{amsmath}
\newcommand{\be}{\begin{equation}}
\newcommand{\ee}{\end{equation}}
\usepackage[centertags]{amsmath}
\usepackage{amsfonts}
\usepackage{caption}
\usepackage{subcaption}
\usepackage{graphicx}
\usepackage{color}

\newcommand{\eqb}{\begin{eqnarray}}
\newcommand{\eqf}{\end{eqnarray}}
\newcommand{\bb}{\begin{equation}}
\newcommand{\beq}{\begin{equation}}
\newcommand{\eeq}{\end{equation}}
\newcommand{\bea}{\begin{eqnarray}}
\newcommand{\eea}{\end{eqnarray}}

\begin{document}
\title{ Vortices in fracton type gauge theories}
\author{G.~Lozano$^a$ and F.~A.~Schaposnik$^b$
\\
~
\\
~
\\
{\normalsize $^a\!$\it Departamento de F\'\i sica, Universidad de Buenos Aires,}\\
{\normalsize $\!$\it IFIBA CONICET, Argentina}\\
{\normalsize $^b\!$\it Departamento de F\'\i sica, Universidad
Nacional de La Plata}\\ {\normalsize\it Instituto de F\'\i sica La Plata-CONICET}\\
{\normalsize\it C.C. 67, 1900 La Plata,
Argentina}
 }

\maketitle
\begin{abstract}
 We consider a vector gauge theory  in $2+1$ dimensions of the type recently proposed by
 Radzihovsky and Hermele \cite{RH} to describe fracton phases of matter. The theory
 has   $U(1) \times U(1)$ vector gauge fields coupled to an additional  vector field with a non conventional  gauge symmetry.  We added to the theory scalar matter in order to break the gauge symmetry. We analyze non trivial configurations by reducing the field equations  to first order self dual (BPS) equations which we solved numerically.   We have found vortex solutions for the gauge fields which in turn generate for the extra vector field  non-trivial configurations that can be associated to  magnetic dipoles.
\end{abstract}
\section{Introduction}
The study of non-trivial solutions in quantum field theories has historically played an essential role in describing non-perturbative phenomena usually linked to topological properties
of theses theories, both in High Energy applications \cite{Shif}  and in   condensed matter systems  \cite{Fradkin}. In the last year,  there has been a growing interest in the study of a new class of quantum states of matter in which quasiparticles called ``fractons'' were introduced in quantum spin-liquid models \cite{Chamon}.
Afterwards,  topological
quantum order was studied  in  Majorana fermion models in which
 only  composites of such elementary excitations
were free to move in certain directions.
Later on, a connection in the low energy limit between fracton phases and  tensor gauge theories was studied  in ref.\ \cite{Pret}. Since then,
interest in the subject grew in various directions of  condensed matter and quantum field theories physics including studies on gravity and   elasticity areas (for reviews see \cite{Nan}-\cite{PretReview}  and references therein).

More recently Radzihovsky and Hermele (RH) have considered a description of fracton phases in $2+1$ dimension in  terms of gauge vector fields \cite{RH}.
The model discussed by these authors consists of   $U(1)\times U(1)$ (conventional) vector gauge fields  coupled to a an additional vector field in such a way that the resulting Lagrangian is invariant under a deformed gauge  transformation.

In this work we will consider a theory where the $U(1) \times U(1)$ vector gauge fields in the RH model  are minimally  coupled to scalar matter implementing the Higgs mechanism. We will show that also this model having an additional vector field  has magnetic like vortex solutions of the Nielsen-Olesen type, which in turn generate a non trivial configuration for the extra vector field of the model. In addition, proceeding as in the original simpler $U(1)$ case, we will be able to reduce the second order field equations to first order self-dual equations \cite{Bogo}-\cite{dVS}. The well known  Nielsen-Olesen ansatz leads to radial equations that can be solved numerically. The solution corresponds to stable vortex magnetic fields associated to the $U(1)\times U(1)$ gauge field sector and an additional magnetic field associated to the extra vector field.

\section{The model}
We shall consider a  $d = 2+1$ dimensional $U(1)\times U(1)$ gauge theory with gauge fields $A_i^a, A_0^a$  with  $i=1,2$ spatial  and $a=1,2$    ``flavor'' indices. There is also an additional vector field $(V_0,V_i)$. The corresponding Lagrangian  density $L_G$ is the one introduced in  \cite{RH} (without external sources),
\begin{equation}
L_G =-\sum_a\frac{1}{4}\,F_{\mu\nu}^a F^{a \mu\nu} + \frac12(\partial_t V_1  +\partial_1 V_0 - A_0^{(1)})^2+
\frac12(\partial_t V_2  +\partial_2 V_0 - A_0^{(2)})^2
  - \frac12 (\epsilon_{ij}\partial_i V_j+ {\cal A} )^2
  \label{LG}
\end{equation}
We assume the standard summation convention for space-time indices with a metric $(-++)$ but we write explicitly sums involving
flavor indices.
Here
   $F_{ij}=\partial_{i} A_{j} - \partial_{j} A_{i}$ and
\be
{\cal A} =\sum_a \epsilon_{ia}A_{i}^a  \label{calAA}
\ee

We will also introduce scalar matter minimally coupled to the  fields $A_\mu^a$ together with a scalar potential to implement gauge symmetry breaking and the Higgs mechanism
\be
L_S = -\sum_a D_\mu \phi^{a \dagger}  D^\mu \phi^a  - V[\phi^a]  \label{LS}
\ee
with the  covariant derivatives ($\mu = 0,1,2$) defined as
\be
D_\mu \phi^a  = (\partial_\mu - ig^a A_\mu^a) \phi^a
\ee
and
\be
V[\phi^a] =\sum_a {\lambda^a} ( |\phi^a|^2 - (\eta^a)^2 )^2
\ee
In principle the potential  could include a mixing $ \tilde{\lambda} |\phi^{(1)}|^2||\phi^{(2)}|^2$ but for simplicity we will assume $\tilde{\lambda} =0$

The total Lagrangian density L is then given by
\be
L=L_G+L_S
\label{2}
\ee
The theory is invariant under ``deformed'' gauge transformations \cite{RH},
\be
\begin{array}{ccc}
\vec A^a_i \to A^a_i + \partial_i \alpha^a & A_0^a \to A_0^a + \partial_{0} \alpha^a,\\   &    \\
V_i \to V_i  + \partial_i\beta - \alpha_i   & V_0 = V_0 - \partial_{0} \beta,
\end{array}
\label{enlarged}
 \ee
 together with
 \be
 \phi^a \to \exp(ig^a\alpha^a) \phi^a
 \ee

In this work we will be interested only in static, purely magnetic configurations, so that the energy density can be written as
\be E =  \sum_a \frac{1}{4}\,F_{ij}^a
F_{ij}^a  +\frac12 (\epsilon_{ij}\partial_i V_j+ {\cal A} )^2+\sum_a(D_i \Phi^a)^{\dagger}\, (D_ i\Phi^a) + V[\phi^a]
\ee
Euler-Lagrange equations are then
\bea
\partial_i F_{ik}^a&=&\epsilon_{ka}(\epsilon_{ij}\partial_i V_j+ {\cal A} )+ig (\phi^a D_k\phi^{a^\dagger}-\phi^{a \dagger}D_k\phi)  \\
D_k D_k \phi^a &=&-\frac{\delta V}{\delta \phi^{a\dagger}} \\
\epsilon_{ij} \partial_k \partial_i V_j  &=&  \partial _k {\cal A}
\eea
Instead of solving these second order field equations, we shall follow the standard
Bogomolny procedure \cite{Bogo} and we  rewrite the energy density as
\begin{align}
&E\!=  \sum_a \left( \vphantom{\sum_a }\frac{1}{2} |D_i \Phi^a - i \gamma^a
\varepsilon_{ij} D_j\Phi^a|^2 +
\frac{1}{4} \left( F_{ij}^a -
\gamma^a g^a \varepsilon_{ij} (\phi^a \phi^{a\dagger}  -
(\eta^a)^2) \right)^2   \right. \nonumber \\
&+\!\left.
\frac12 (\epsilon_{ij}\partial_i V_j+ {\cal A})^2 +
\sum_a\left(\lambda^a - \frac{(g^a)^2}{2} \right)  (\Phi^a\, \Phi^{a\dagger}  -  (\eta^a)^2)^2-
\gamma^a  \frac{g^a}{2}
(\eta^a)^2
\varepsilon_{ij}  F_{ij}^a \right) \nonumber \\
\label{ener2}
\end{align}
where $\gamma^a=\pm $1 and we have discarded total derivatives which vanish after integration for appropriate boundary conditions (in this case we require finite energy in $R^{(2)}$ which implies vanishing of the scalar covariant derivatives at infinity).
So, if
\be
\lambda^a = \frac{(g^a)^2}{2}
\label{point}
\ee
the minimal value of the energy ${\cal E}$
\be
{\cal E} = \int d^2x E
\ee
is reached when the three squared terms in eq.\eqref{ener2} vanish
\bea
&&D_i \Phi^a - i \gamma^a \,
\varepsilon_{ij}\, D_j\Phi^a=0  \label{uno}\\
&& F_{ij}^a -
\gamma^a\, g^a\, \varepsilon_{ij} (\Phi^a\, \Phi^{a\dagger}  - \,
(\eta^a)^2) =0 \label{dos}
\\
&&\epsilon_{ij}\partial_i V_j+ {\cal A}=0 \label{tres}
\eea
If eqs. \eqref{uno}-\eqref{tres} are satisfied
the energy ${\cal E}$ is
\be
{\cal E} = \int d^2 x E =- \sum_a\gamma_a  g_a(\eta^a)^2 \int d^2x B^a = 2\pi  \sum_ a(\eta^a)^2 |m^a|
\ee
where $m^a$ is the winding number associated to the quantized magnetic flux .

Now, the simplicity and convenience of the self dual equations are apparent. Equations for $(A_i^{(1)},\Phi^{(1)})$ a $(A_i^{(2)}, \Phi^{(2)})$ are first order and decoupled. After solving them, we can use
$A_i^{(1)},A_i^{(2)} $ as sources for $V_i$. On the the hand, the energy can be calculated explicitly and their stability is ensured because they satisfy the Bogomolny bound. The
self-dual equations are valid only when the relation Eq. \eqref{point}  is valid. It is simple to see that this relation implies the equality between vector and scalar masses of the theory. It is also well established the connection between the existence of self dual equations and $N=2$ supersymmetry for several models \cite{dVS}-\cite{ENS}. In the original
Ginzburg-Landau theory of superconductivity (which has a single $U(1)$ sector) relation Eq. \ref{point} signals the boundary between Type I and Type II superconductors.

We will look
 for axially symmetric configurations
for $(A_i^{(1)},\Phi^{(1)})$ and $(A_i^{(2)},\Phi^{(2)})$, so we make the following
ansatz in polar coordinates $(r,\varphi)$
\bea
A^a_\varphi&=&-A_x^a r\sin\varphi+ A_y^a r \cos\varphi =\frac{1}{g_a} a_\varphi(r) \\
A^a_r&=&A_x^a \cos\varphi+ A_y^a \sin\varphi=0 \\
\Phi^a&=& \eta^a f^a(r) e^{i m_a \varphi}
\eea

Then, the first two equations become
\bea
\partial_r f^a&=&-\frac{\gamma^a}{r}(m^a-a^a_\varphi)f^a \\
\frac{1}{r}\partial_r a_\varphi^a&=& (\eta^a)^2 \gamma^a (g^2) ^a ((f^a)^2-1)
\eea
Finite energy requires the following boundary conditions
\bea
a^a_\varphi(0)&=&0\,,  \hspace{1.5 cm} a^a_\varphi(\infty)=m^a \nonumber \\
f^a(0)&=&0\,, \hspace{1.5 cm} f^a(\infty)=1
\eea
It is easy to check that consistency requires
$
{\gamma^a}/{m^a} < 0
$.
It will also be convenient to redefine
\be
\rho=|g^1 \eta| r \,\,\,\,\,\,\, \tilde{a}_\varphi^a= {a}_\varphi^a-m^a
\ee
then
\bea
\partial_\rho f^{(1)}&=&\frac{\gamma^{(1)}}{\rho}(\tilde{a}^{(1)}_\varphi)f^{(1)}    \,\,\, \,\,\, \frac{1}{\rho}\partial_\rho \tilde{a}_\varphi^{(1)}=  \gamma^{(1)}  ((f^{(1)})^2-1) )\\
\partial_\rho f^{(2)}&=&\frac{\gamma^{(2)}}{\rho}(\tilde{a}^{(2)}_\varphi)f^{(2)}    \,\,\,   \,\,\, \frac{1}{\rho}\partial_\rho \tilde{a}_\varphi^{(2)}= \gamma^{(2)} \delta^2 ((f^{(2)})^2-1) )
\eea
where $
\delta=\frac{g^{(2)} \eta^{(2)}}{g^{(1)} \eta^{(1)}}
$
and
\bea
\tilde{a}^a_\varphi(0)&=&m^a\,, \hspace{1.5 cm} \tilde{a}^a_\varphi(\infty)=0 \nonumber \\
f^a(0)&=&0 \,, \hspace{1.5 cm} f^a(\infty)=1
\eea
One can then show that
\bea
&&\partial^2_{\rho^2}\tilde{a}^{(1)}_\varphi-
\frac{1}{\rho}\partial_\rho \tilde{a}^{(1)}_\varphi
(1+2\gamma^a\tilde{a}^{(1)}_\varphi)-2\tilde{a}^{(1)}_\varphi=0
\\
&&\partial^2_{\rho^2}\tilde{a}^{(2)}_\varphi-
\frac{1}{\rho}\partial_\rho \tilde{a}^{(2)}_\varphi
(1+2\gamma^a\tilde{a}^{(2)}_\varphi)-2\delta^2\tilde{a}^{(2)}_\varphi=0
\eea
It is obvious that, if $m^{(1)}=m^{(2)}$ then $\tilde{a}^{(2)}(\rho^{*})=\tilde{a}^{(1)}(\rho)$
with
$\rho^{*}=\delta\rho$.

The equation \eqref{tres} for $V_i$ can be now written in terms $a_\varphi^{(1)}$ and $a_\varphi^{(2)}$,
\be
\tilde{B}\equiv\epsilon_{ij}\partial_i V_j=-A_x^{(2)}+A_y^{(1)}=\frac{a_\varphi^{(1)} \cos\varphi +a_\varphi^{(2)} \sin\varphi}{r}
\label{tilde}
\ee
so that once we have solved the equations  for $a_\varphi^a$, we can easily  obtain the solution for $\tilde B$.

\begin{figure}[t]
    \centering
    \includegraphics[width=17cm]{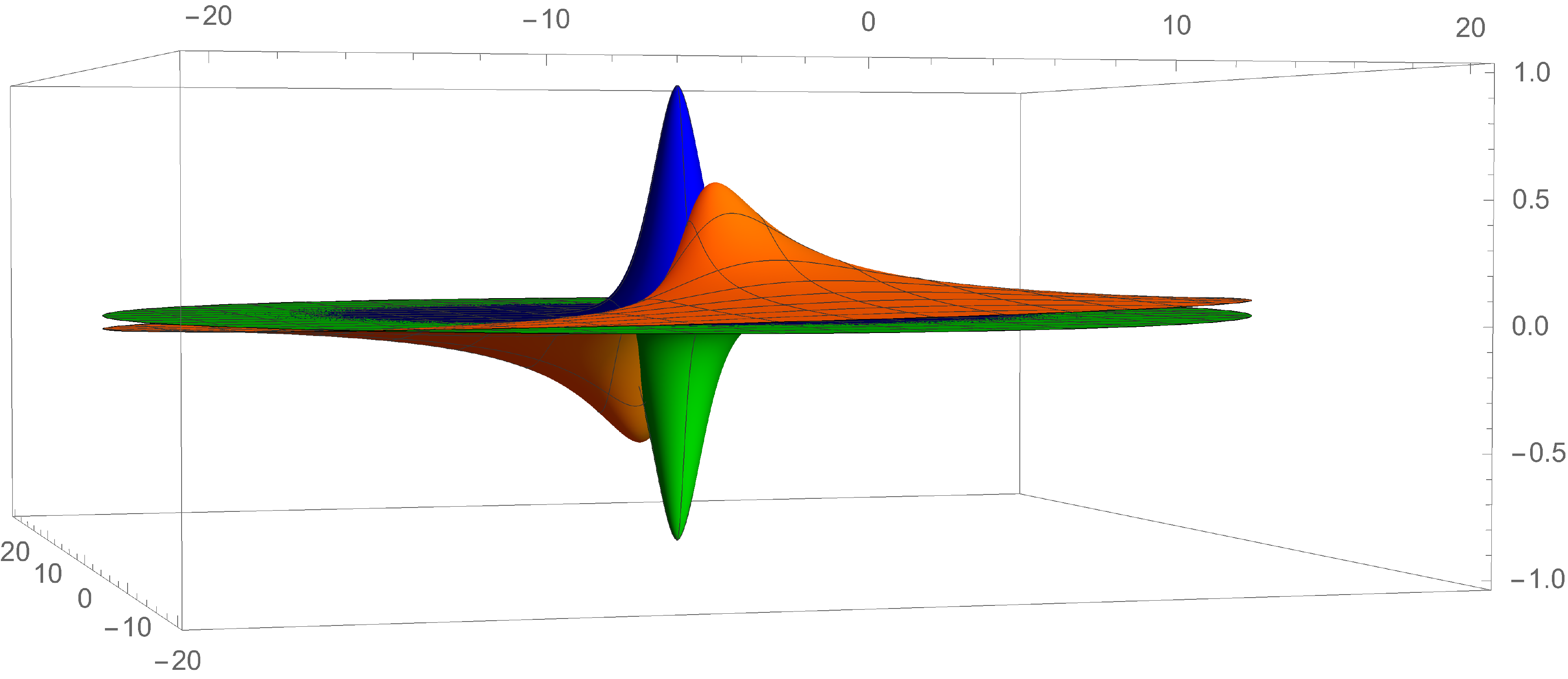}
    \caption{We show the magnetic field $B^1$ associated to a vortex with winding number
    $m^{(1)}=1$, (blue), $m^{(2)}=-1$ (green)  and the magnetic field associated to the field $a_i$ (orange). Parameters have been chosen so that  $g^a = 1$ and $\eta^a=1$ for $a=1,2$.}
    \label{dip}
\end{figure}

We have found numerical solutions of
Eqs. (29)-(31) by using a relaxation method. We have analyzed different topological sectors with different winding numbers  $(m^{(1)},m^{(2)})$ and fluxes $B^a$
\be
\Phi^a =\int d^2x B^a=\int d^2x F_{12}^a= \frac{2\pi}{g^a}m^a
\ee
We show in  Fig.\ref{dip}  a solution for the case in which the topological numbers $m^{(1)} = 1$ and $m^{(2)} = -1$,  and where for simplicity we have set $g^{(1)}= g^{(2)}=1$ and $\eta^{(1)} =\eta^{(2)} $. The upper  peak (in blue) corresponds to the magnetic field associated with the vortex with winding number $m^{(1)}=1$, and the lowest one to the magnetic field of the vortex with $m^{(2)}=-1$.
In the same plot (in orange) we show the  $\tilde B$ field defined in Eq. \eqref{tilde}, which  present a double peak structure. We remark that the particular (mirror) symmetry of the figures originates from our choices for $\eta^a$ and $m^a$ but more generic cases can be considered without additional computational effort.

\begin{figure}[t]
\centering
  \begin{subfigure}{7cm}
    \centering\includegraphics[width=6.5cm]{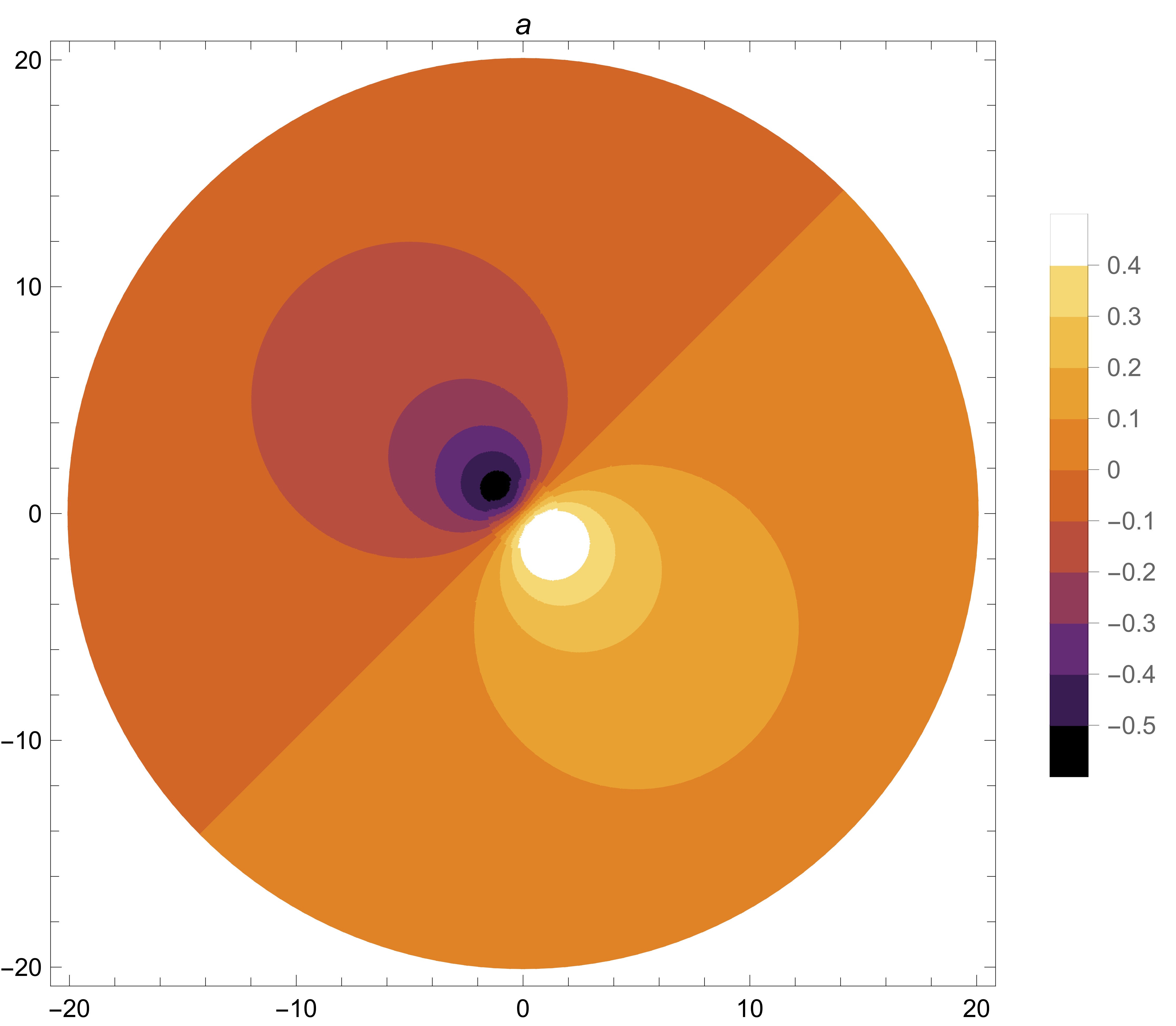}
    \caption{sector (1,-1)}
  \end{subfigure}
  \begin{subfigure}{7cm}
    \centering\includegraphics[width=6.5cm]{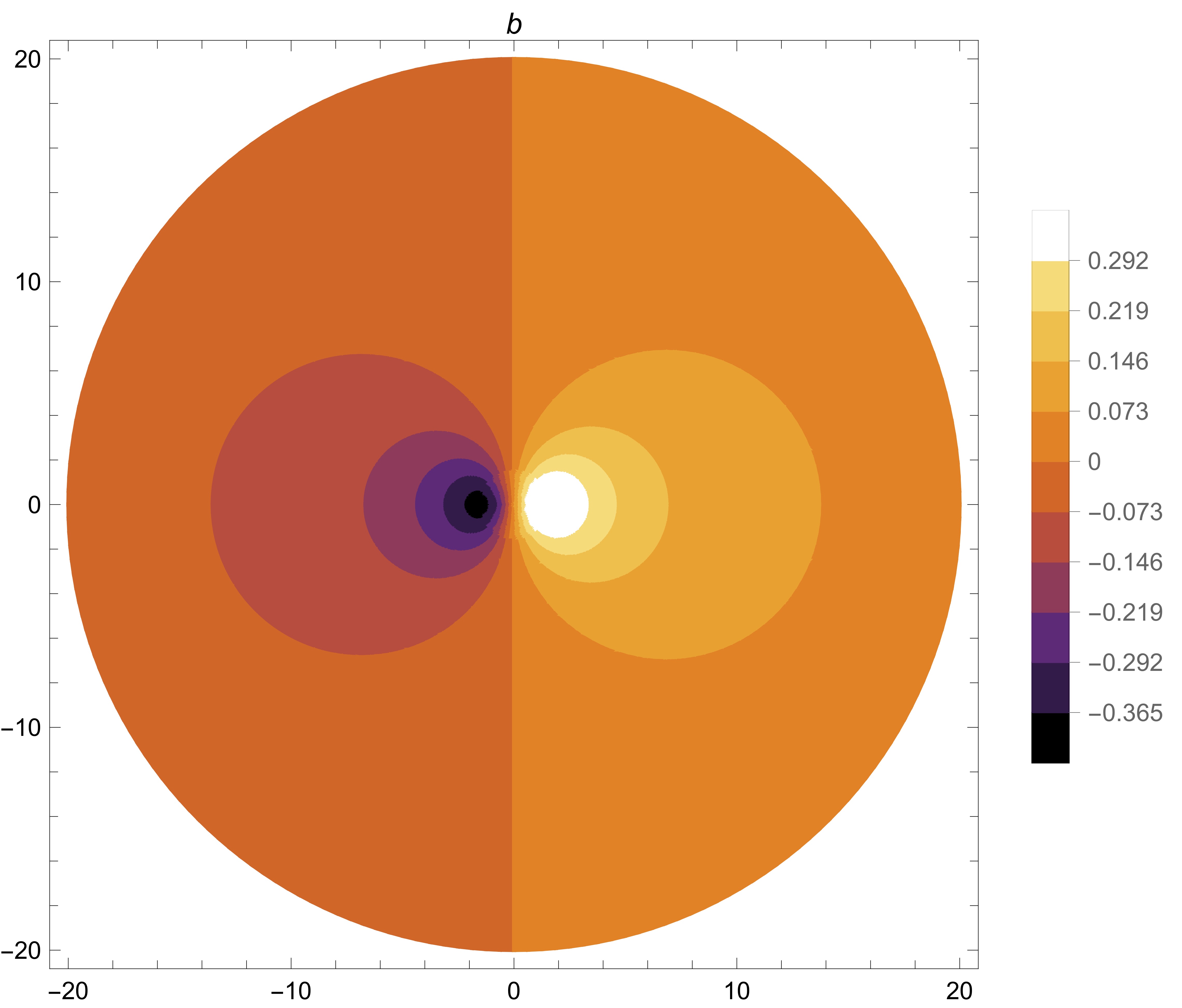}
    \caption{sector (1,0)}
  \end{subfigure}
      \caption{Contour plots for the magnetic field associated to the field $\tilde B[a_i]$ for different winding numbers $(m^{(1)}, m^{(2)})$. Parameters have been chosen so that  $g^a = 1$ and $\eta^a=1$ . Panel $(a)$ corresponds to contour plots for  sector $(1,-1)$ while panel $(b)$ to sector $(1,0)$ .}
    \label{B}
\end{figure}

\begin{figure}[t]
 \centering
 \includegraphics[width=5cm]{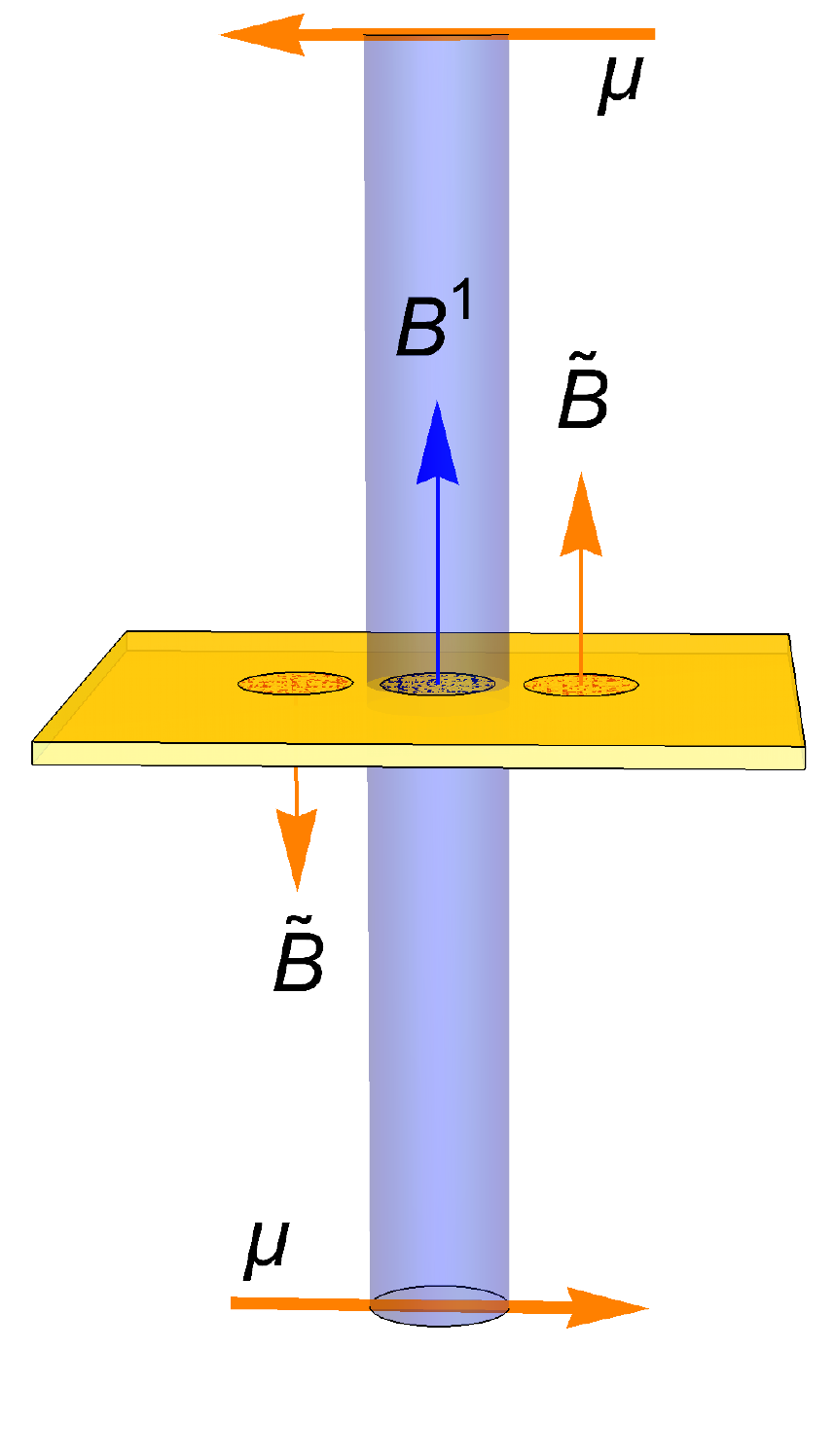}
    \caption{We represent in three spatial dimensions the magnetic field associated to a $B^{(1)}$ vortex with
    $m^{(1)}=1$, (light blue)   and the $\tilde B$ field associated to   $a_i$ (orange). Orange arrows $\mu$ represent the two effective magnetic dipoles, in this case in the $x$ direction.}
    \label{A}
\end{figure}

Notice that the sources of this generalized magnetic field are the vector potentials of the $U(1)\times U(1)$ sector via the term $\sum_a \epsilon_{ia}A_{i}^a$. Thus, both $U(1)$  gauge fields contribute to the $\tilde B$ field. Nevertheless,  it is enough to have only  one of these gauge fields different from zero to produce a non-zero $\tilde B$ field. Indeed in Fig  \ref{B}   we display  contour plots of the $\tilde B$ field for two different choices of  gauge fields of the  $U(1)\times U(1)$ sector. Panel $(a)$ corresponds to the contour plot of $\tilde B$ associated to the Fig \ref{A}, this is $(m^{(1)},m^{(2)})=(1,-1)$. Panel $(b)$, corresponds to a contour plot of $\tilde B$ for the case $(m^{(1)},m^{(2)})=(1,0)$, where only the $A^{(1)}$ acts as a source for $\tilde B$. Notice that not only the intensity of the field changes depending on the choice of $m^a$ but also figure in the panel (b)  is rotated with respect to the one in panel $(a)$.
\

Looking in more detail to panel $(b)$ in Fig \ref{B}, the contour plot looks qualitatively very similar to those of the magnetic field produced by a magnetic dipole placed outside the $(x,y)$ plane, at a certain height in a $z$ axis in three spatial dimensions,     as represented schematically in Fig \ref{A} for the $A^{(1)}$ field. In this figure we display the $B^{(1)}$ magnetic field tube (in light blue), and the  two effective magnetic dipoles $\vec \mu$  (represented as  orange arrows) associated to the  $\tilde{B}$  field.
    Notice that the direction of the dipole is correlated with the flavor of  the gauge field ($a=1$ in this case). Had we chosen, the other flavor $a=2$, the orientation of the dipole would be different. In fact, panel $(b)$ of Fig \ref{B}. results from the superposition of these cases.
\begin{figure}[t]
 \centering
 \includegraphics[width=15cm]{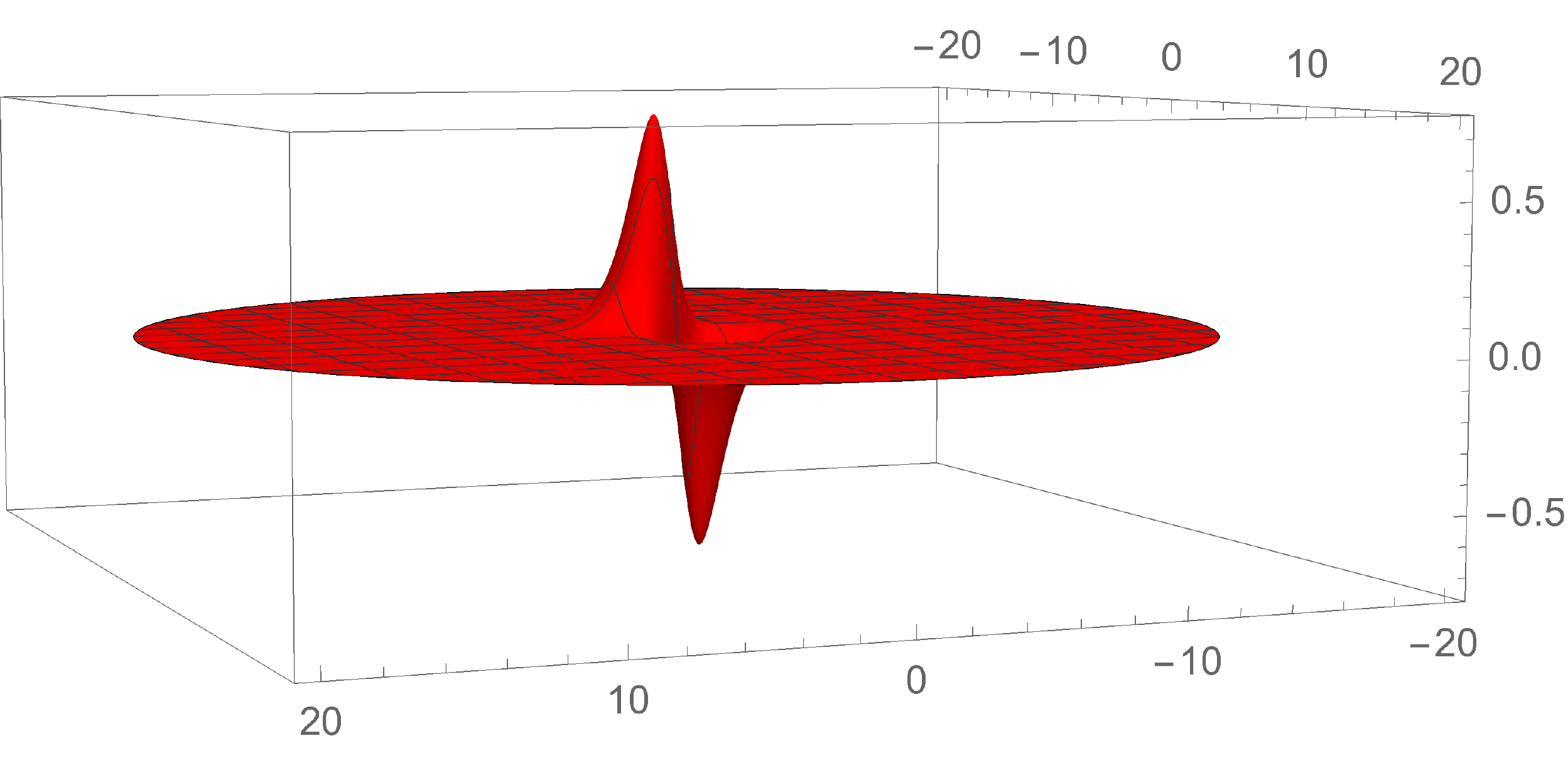}
    \caption{ Plot of the density $j_0^m$ for the $(1,-1)$ sector . Parameters have been chosen so that  $g^a = 1$ and $\eta^a=1$ }
    \label{D}
\end{figure}
Associated to the generalized gauge transformation of the $V_i$ field, a conserved (and gauge invariant) density $j_0^m$ was identified  in
 Ref \cite{RH},,
\be
j_0^m =\epsilon_{ij}\partial_i V_j+ {\cal A}- r_i \epsilon_{ij} J^m_{0j}
\ee
with
\be
J^m_{0j} = \sum_l \epsilon_{jl} B^{(l)}
 \ee
 In our   case $j_0^m$ reduces to
\be
j_0^m=x B^{(1)}+y B^{(2)}
\label{densidad}
\ee
We show in Fig. \ref{D} a plot of this density $j_0^m$ for the case in in which $(m^{(1)},m^{(2)})=(1,-1)$. Similar plots can be obtained for the different sectors.

The  Lagrangian \eqref{LG} proposed in \cite{RH}, when coupled to  appropriate currents leads to Gauss law of a symmetric tensor gauge  theory coupled to an external {\em electric} charge  which encodes conservation not only of such charge but also conservation of an electric dipole moment \cite{PretReview}. In the present  work, we have shown that  the model described by Lagrangian \eqref{2},  presents an interesting magnetics sector  where in addition to the Nielsen-Olesen type vortex solutions typical of the standard $U(1)$ Higgs models,  the coupling between the gauge fields and the vector field $V_i$ gives rise to  additional {\em magnetic} fields which are qualitatively similar to those produced by  an effective   magnetic dipole as reflected by the conserved  density $j_0^m$.

A relevant question that arises is what could it be the role of this kind of structures in fracton models.  Even a more interesting situation could  be expected if  in addition to  Maxwell term considered here, a Chern-Simons term is added .  It is well-known that in the presence of Chern Simons term in  the standard case  vortices that carry both, electric and magnetic charge  are present \cite{Paul}-\cite{deVega2}. It is also well known that BPS equations can be found for conveniently tuned models both in the relativistic and non-relativistic cases \cite{Hong}-\cite{J2}. We expected that the analysis presented here can be also applied to this case. We hope to report on this issue soon. 

\vspace{1.2 cm}

\noindent{\bf{Acknowledgments:}}
F.A.S. is financially supported by PIP-CONICET (PIP688),
and UNLP grants X910.  G.S.L. is financially supported by PICT2016-1212, PIP 11220150100653CO, Conicet, y UBACYT  20020170100496BA.

\end{document}